\documentclass[aps,showpacs,superscriptaddress,amsfonts,twocolumn,prl]{revtex4}

\usepackage{graphicx}
\usepackage{array}
\usepackage{hyperref}
\usepackage{amsmath,amssymb}
\usepackage{float}
\usepackage{amsfonts}
\usepackage{ar}
\usepackage{xcolor}
\usepackage{physics}
\usepackage{soul}
\usepackage{siunitx}

\begin{document}

\title{Microbial narrow-escape is facilitated by wall interactions}

\author{Mathieu Souzy $^*$}
\affiliation{ Physics of Fluids, University of Twente, The Netherlands}
\author{Antoine Allard $^*$}
\affiliation{Department of Physics, University of Warwick, Gibbet Hill Road, Coventry CV4 7AL, UK}
\author{Jean-Fran\c{c}ois Louf}
\affiliation{Department of Chemical Engineering, Auburn University, Auburn, AL 36849, USA}
\author{Matteo Contino}
\noaffiliation
\author{Idan Tuval}
\affiliation{Mediterranean Institute for Advanced Studies, IMEDEA, UIB-CSIC, Esporles, 07190, Spain}
\affiliation{Physics Department, Universitat de les Illes Balears, 07122, Palma de Mallorca, Spain}
\author{Marco Polin}
\affiliation{Department of Physics, University of Warwick, Gibbet Hill Road, Coventry CV4 7AL, UK}
\affiliation{Mediterranean Institute for Advanced Studies, IMEDEA, UIB-CSIC, Esporles, 07190, Spain}
\affiliation{Physics Department, Universitat de les Illes Balears, 07122, Palma de Mallorca, Spain}

\date{\today}

\begin{abstract}

Cells have evolved efficient strategies to probe their surroundings and navigate through complex environments. From metastatic spread in the body to swimming cells in porous materials, escape through narrow constrictions - a key component of any structured environment connecting isolated micro-domains - is one ubiquitous and crucial aspect of cell exploration. Here, using the model microalgae \textit{Chlamydomonas reinhardtii}, we combine experiments and simulations to achieve a tractable realization of the classical Brownian narrow escape problem in the context of active confined matter. Our results differ from those expected for Brownian particles or leaking chaotic billiards and demonstrate that cell-wall interactions substantially modify escape rates and, under generic conditions, expedite spread dynamics.

\end{abstract}

\maketitle
\def\thefootnote{*}\footnotetext{These authors contributed equally to this work}\def\thefootnote{\arabic{footnote}}

From cytoskeletal dynamics and bacterial swarming to flocks of birds and human crowds, collections of interacting self-propelling agents can display unexpected emergent properties as a consequence of being far-from-equilibrium. Understanding these within the framework of a general theory is a challenge which is currently driving the rapidly growing area of active matter physics \cite{Marchetti2013}. At the micro-scale, paradigmatic examples include cell aggregates and microorganisms, where motility underlies numerous critical biological processes such as infection and biofilm formation by bacteria, cancer metastasis, tissue repair and wound healing, and complex morphogenesis of new tissues and organs \cite{Hartmann2019,Needleman2017,Collinet2021}. Research on these processes have already provided substantial insight into the biological and physicochemical mechanisms that engender and regulate cellular motility in open uniform domains. However, cells often inhabit complex and heterogeneous three-dimensional environments like gels and tissues in the body or soils and sediments in the environment, that impose additional geometric constraints, mechanical cues, and external stimuli such as chemical gradients and fluid flows. These factors fundamentally alter cellular motility, hindering or promoting active transport in unexpected ways, and giving rise to fascinating behaviours such as directed cell migration and large-scale coordination \cite{Bricard2013,Rein2016}.

\begin{figure}[hb]
	\includegraphics[width=0.95\columnwidth]{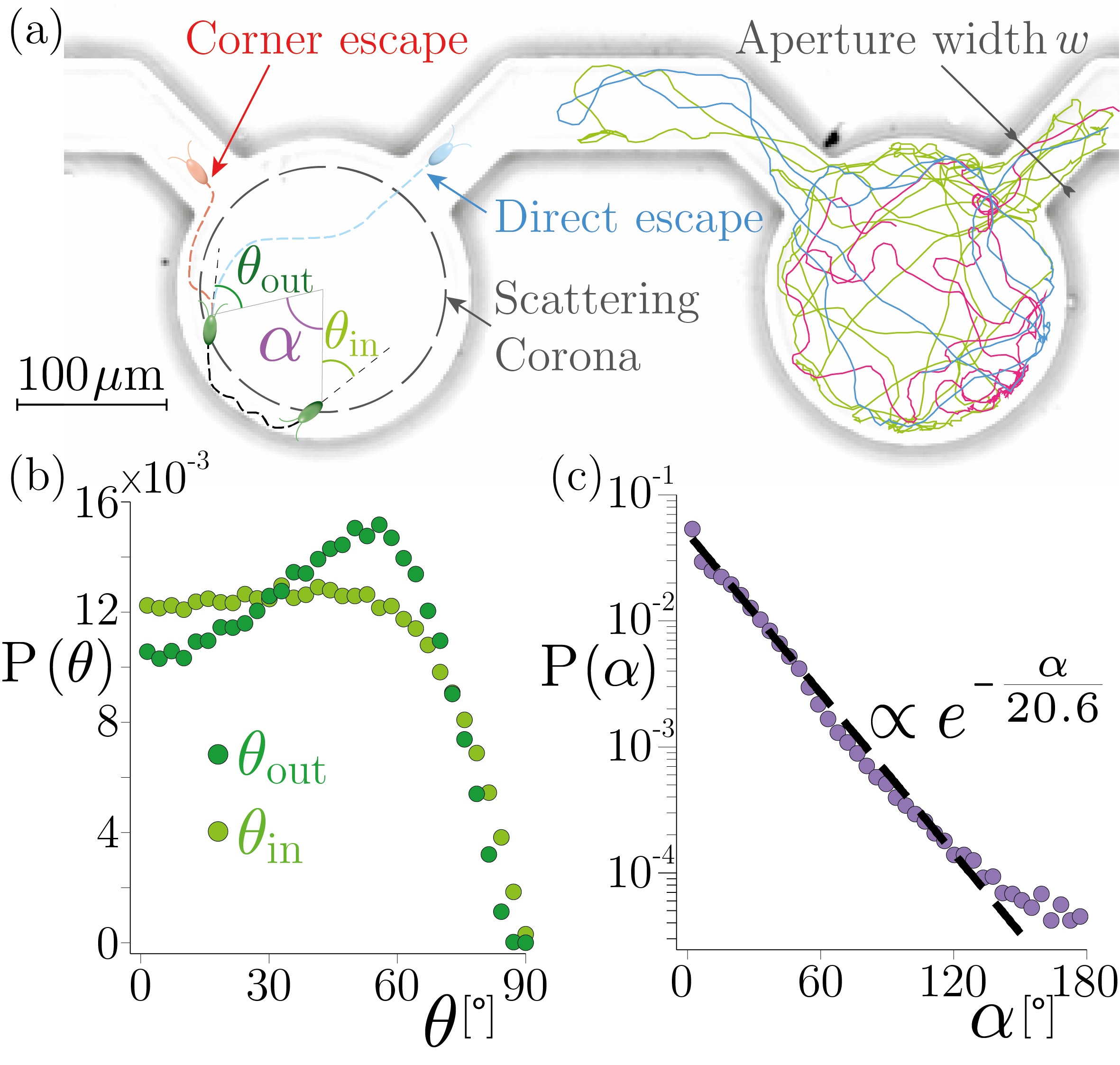}
	\caption{\label{fig:Schematic} \textbf{Tracking of CR escape through narrow apertures.} (a) Left pool: schematic of corner and direct escapes. Right pool: typical trajectories of N = 3 algae travelling through and escaping a pool of aperture width $w$ (b) Distribution of $\theta_{in}$ and $\theta_{out}$ (obtained over $\sim 2.4\cdot10^5$ events). (c) Distribution of sliding angle $\alpha$ (obtained over $\sim 2\cdot10^5$ events). The fit is an exponential function with a characteristic angle	$\bar{\alpha}=20.6^{\circ}$.}
\end{figure}

Current research efforts in this area aim at expanding our understanding of the dynamics of active matter under these constraints to incorporate the complexity of confined, crowded and structured environments \cite{Bechinger2016}. In many cases, such complex environments challenge preconceived ideas based on studies of either active matter in uniform domains or passive (thermally equilibrated) matter in complex environments. Examples include persistent directional motion in structured environment under external (e.g.,  chemotactic) cues \cite{Phan2020}, the effect of disordered fluid flows \cite{DeAnna2021}, coordinated cell migration \cite{Bruckner2021, Renkawitz2019} or cell hopping and trapping through porous media \cite{Bhattacharjee2019b}. Further progress will stem from the use of accessible experimental realizations.

Using the model unicellular eukaryote \textit{Chlamydomonas reindhardtii} (CR) \cite{Harris2001, Harris2009a, Goldstein2015}, we present here experimental and numerical work that captures the essential features of active particles escaping confinement through narrow apertures, a key component of any structured environment representing the minimal `bridge unit' that connects isolated spatial components or micro-domains. For motile bacteria, dynamics through narrow channels gives rise to bistable density oscillations and stable along-axis swimming \cite{Paoluzzi2015,Vizsnyiczai2020}. For microalgae, the escape dynamics through narrow apertures resembles an out-of-equilibrium version of the classical Brownian narrow escape problem so ubiquitous in physics, chemistry, and biology. This beautiful and mathematically intricate problem, first studied by Lord Rayleigh in the context of acoustics \cite{Rayleigh1878}, has  recently received much attention as the mean escape time controls the rates of fundamental molecular processes (e.g., from mRNA escaping through nucleus pores in the cell \cite{Berg1981} to signalling in dendritic spines \cite{Harris1988, Holcman2005}). Instead, for non-Brownian particles following purely ballistic motion, the escape dynamics is captured by the theory of leaking chaotic systems \cite{Altmann2013}, with an exponential decay in particle number only expected for chaotic dynamics, while so-called deterministic `billiards' give rise to a 1/t decay \cite{Bauer1990, Krieger2016, Spagnolie2017}. 

Here, we experimentally realize a minimal version of the narrow escape problem in the context of confined active particles. While recent theoretical works have numerically addressed the effect of swimming characteristics (e.g., persistence length as compared to the domain size \cite{Caprini2019, Paoluzzi2020}) or the role of collective interactions \cite{Olsen2020} for the escape dynamics, we instead focus on the relevance of cell-wall interactions for determining the escape rate. In particular, we tackle the contribution of  recently described cell sliding on curved surfaces \cite{Contino2015, Sipos2015, Ostapenko2018, Thery2021, Cammann2021} and find that it dominates the escape process for aperture sizes comparable to the size of individual cells.

\textit{Chlamydomonas reinhardtii} wild type strain CC125 (CR) was grown in tris-acetate-phosphate medium at \SI{21}{\celsius} under periodic fluorescent illumination (\SI{14}{\hour}/\SI{10}{\hour}), as previously described \cite{Mathijssen2018}. Culture was harvested in the exponential phase ($10^6$ cells/mL). CR were loaded into a polydimethylsiloxane (PDMS) microfluidic device, previously passivated with a 0.5\% w/v Pluronic F-127 solution, made of a set of quasi two-dimensional cylindrical pools of radius $R_{\rm pool} \sim \SI{100}{\micro\meter}$ and height $h\sim \SI{20}{\micro\meter}$ interconnected by narrow channels of different aperture sizes $w$, in the range of $10$ to \SI{75}{\micro\meter} (Fig.~\ref{fig:Schematic}a). Each pool presents two exits located on the pool perimeter at an angular distance of $90^{\circ}$ from each other. Cells were recorded at 20 fps, with a spatial resolution of \SI{1.85}{\micro\meter\per pixel} to gather accurate statistics on algae escape trajectories over long periods of time (up to tens of minutes). Tracking was performed using an in-house developed Kalman filter based algorithm, particularly efficient for tracking relatively dense flow of moving objects \cite{Kalman1960,Souzy2020}.

Typical algae trajectories are shown on Figure \ref{fig:Schematic}a (and on supplementary movies showing single-cell tracking \cite{ref-Movie1,ref-Movie2}) from which we fully characterize free motility parameters (swimming speed, $v=70 \pm50\,\mu\rm{m.s}^{-1}$ and rotational diffusivity in the bulk, $D=\SI{1.1}{\square\radian\per\second}$). It is apparent that cells strongly interact with pool boundaries and tend to transiently swim aligned to the pool walls when sufficiently close to them. This interaction is reflected in a net accumulation of algae inside a narrow ($\sim 17\,\mu$m wide) annulus close to the chamber walls \cite{Ostapenko2018}. Following \cite{Contino2015}, we call this interaction area the `scattering corona', and its width approximately corresponds to the cell body size ($ r_{\rm{algae}}\sim\SI{5}{\micro\meter}$) extended by flagella length ($\ell_{\rm{flagella}} = \SI{12}{\micro\meter}$). Cell-wall interactions can be simply described by the incoming and outgoing angle distributions with respect to a direction orthogonal to the wall ($\theta_{\rm in}$ and $\theta_{\rm out}$, respectively) when entering or exiting the scattering corona. The distribution of  $\theta_{\rm in}$ is rather constant up to $\theta_{\rm in}\simeq\SI{60}{\degree}$, and sharply decreases to zero when cells arrive tangentially to the corona (Fig.~\ref{fig:Schematic}b). The distribution of $\theta_{\rm out}$ peaks at about \SI{60}{\degree}, which implies that cells tend to exit the corona with a preferred angle after interacting with the wall. In between these two events, cells slide along the pool wall for an angular extent, $\alpha$, well approximated by a simple exponential distribution (Fig.~\ref{fig:Schematic}c). As a consequence, we will heuristically describe wall sliding as a Poisson process with a characteristic angle $\bar{\alpha} \simeq \SI{20.6}{\degree}$.

\begin{figure}[t]
	\includegraphics[width=1.\columnwidth]{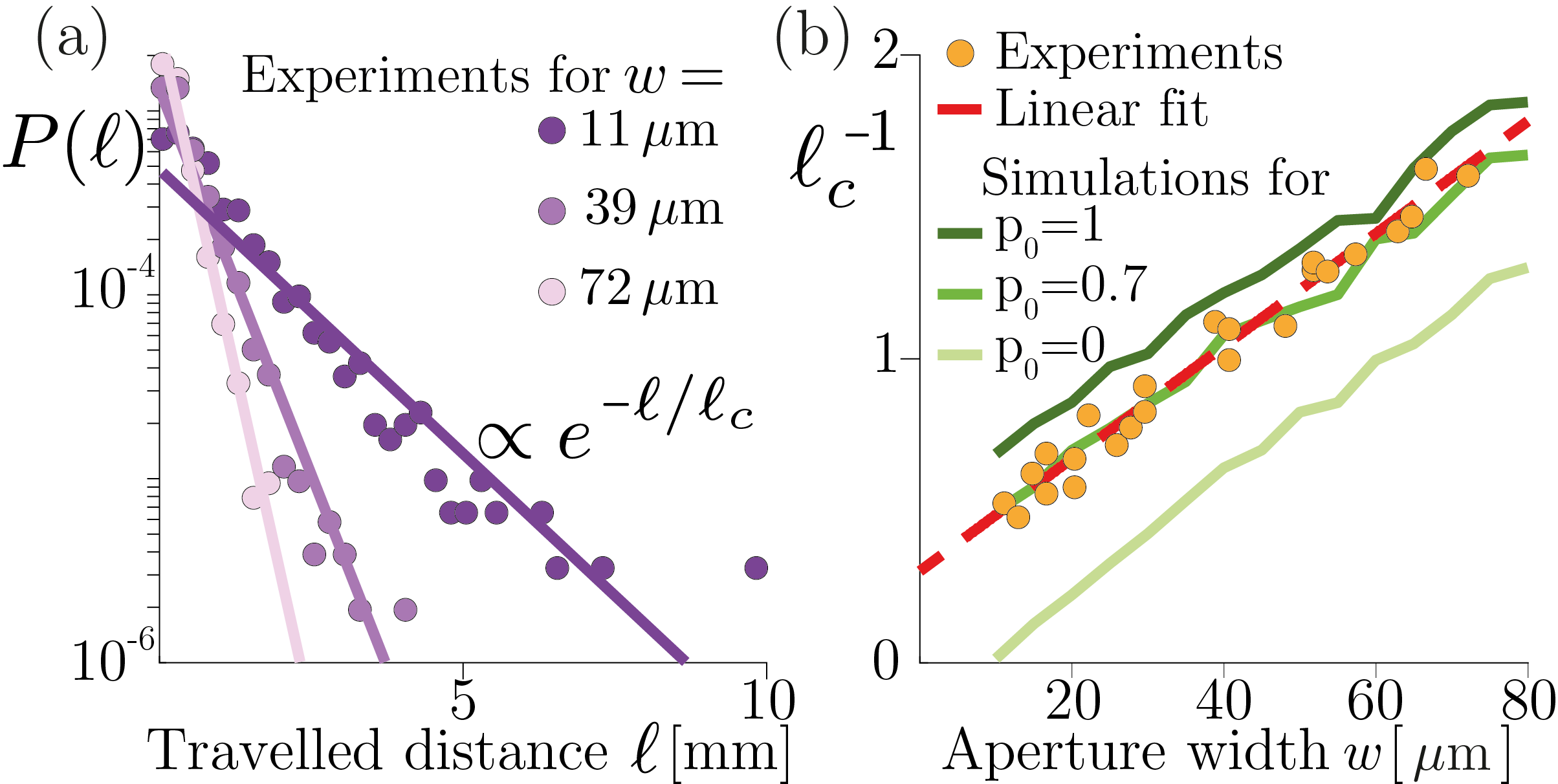}
	\caption{\label{fig:Travelled distance} \textbf{Travelled distance in the pool.} (a) The distribution of travelled distances $\ell$ exponentially decays with a characteristic length $\ell_{\rm c}$. Results are shown here for 3 different aperture widths, with the corresponding exponential fits (lines). (b) The characteristic length $\ell_{\rm c}$ inversely decays with the aperture width. Simulations display the same tendency for various values of $p_0$, the probability to escape after sliding.}
\end{figure}

Turning to the characterization of the narrow escape process, we select trajectories for which individual cells can be followed for the complete duration of trapping events, i.e. from the time they enter a pool through the connecting channels until the time of escape. The measured persistence length of individual trajectories, $\ell_P = v/D_{\theta} \simeq \SI{65}{\micro\meter} \lesssim R_{pool}$, places the dynamics close to the crossover to the diffusive regime described in \cite{Paoluzzi2020}. For all aperture sizes considered, the travelled distance in the pool before escaping is exponentially distributed (Fig.~\ref{fig:Travelled distance}a), as expected for leaking chaotic billiards, with the characteristic length $\ell_{\rm c}$ inversely decreasing with the aperture size (dots, Fig.~\ref{fig:Travelled distance}b) -- cells travel a larger distance before escaping when trapped in pools connected by narrower channels. However, a linear fit to $\ell_{\rm c}^{-1}(w)$ shows a positive non-zero intercept ($0.30 \pm 0.06 ~\mu m^{-1}$) which strikingly contrasts with the solutions to the Brownian narrow escape problem, characterized by the divergence of the mean escape time. Or, in other words, short $\ell_{\rm c}$ are overrepresented in the dynamics (in particular for values of $w$ approaching cell size) as compared to trivial intuition. This hints to an as yet overlooked mechanism facilitating escapes. 

\begin{table}[tb]
	\caption{\label{tab:parameters}List of parameters used for the simulations.}
	\begin{tabular}{|l|l|}
		\colrule
		\textrm{Parameter}&
		\textrm{Value}\\
		\hline\hline
		Radius of the pool & $R_{\rm pool} = \SI{101.6}{\micro\meter}$\\
		Radius of the swimmer & $a = \SI{3.8}{\micro\meter}$\\
		Velocity of the swimmer & $v = \SI{70}{\micro\meter\per\second}$\\
		Rotational diffusion & $D = \SI{1.1}{\square\radian\per\second}$\\
		Sliding angle & $\bar{\alpha} = \SI{20.6}{\degree}$\\
		Scattering angle & $\theta_{\rm out} = \SI{41.5}{\degree}$\\
		\colrule
		Time step & $\Delta t = \SI{0.1}{\second}$\\
		Number of cells & $N_{\rm cells} = \num{e3}$\\
		Number of steps & $N_{\rm steps} = \num{e4}$\\
		\colrule
	\end{tabular}
\end{table}

Indeed, a careful re-examination of all recorded escape events shows that these can be categorized into two distinct classes: \textit{direct escapes}, in which cells directly swim towards the aperture from bulk after a last scattering with distant walls; and \textit{corner escapes}, where cells slide along the curved walls to reach the aperture and eventually leave the pool without leaving the scattering corona (Fig. \ref{fig:Schematic}a). We further explore the relevance of this categorization by building a minimal model of the escape process as follows: we simulate $N_{\rm cell}$ non-interacting circular swimmers of radius $a$ confined to a 2-dimensional circular pool of radius $R_{\rm pool}$ with two openings of width $w$ set at \SI{90}{\degree} apart. Cells' position and orientation are randomly initialized with all parameters (e.g., swimming speed $v$ and rotational diffusivity $D$) extracted from our experimental data and summarized in Table~\ref{tab:parameters}. Over the course of a numerical simulation, if the cell reaches an aperture without touching the wall, then it directly escapes the pool. However, if the finite size swimmer enters in contact with the pool boundary it slides for a sliding angle $\alpha$ randomly picked from the experimentally measured exponential distribution (Fig.~\ref{fig:Schematic}c and Table~\ref{tab:parameters}). If a cell slides for a distance greater than that needed to reach the closest aperture, it escapes the pool with a probability $p_0$ through a corner escape. Otherwise, it remains in the pool and it is scattered back from the corona towards the bulk with an angle $\theta_{\rm out}$ corresponding to the mean experimental scattering angle (using instead the peak value for $\theta_{\rm out}$ only has a minor quantitative effect on the presented results).

\begin{figure}[tb]
	\includegraphics[width=1.\columnwidth]{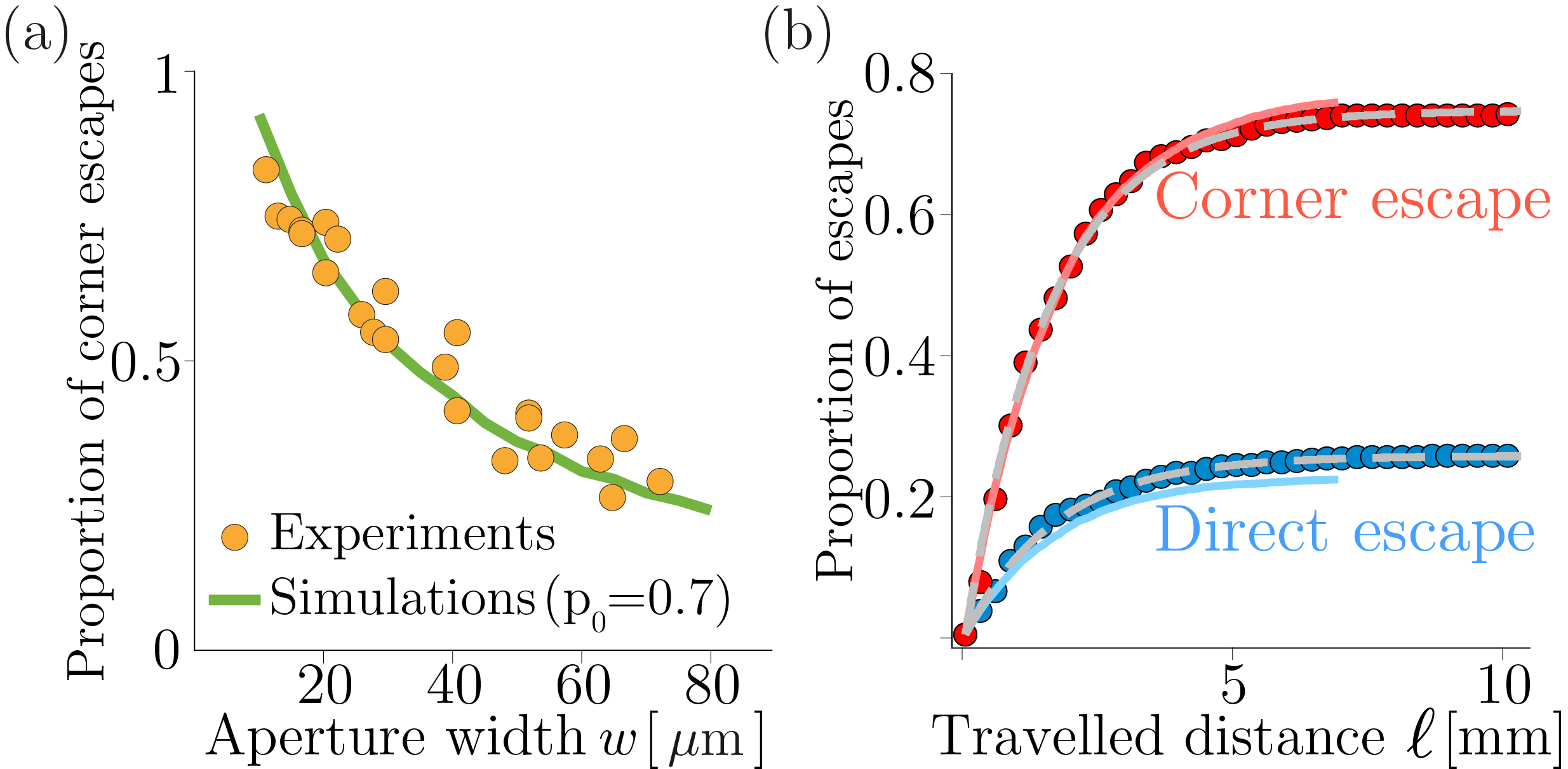}
	\caption{\label{fig:Corner escapes} \textbf{Corner escapes.} (a) Ratio of corner escapes relative to the total number of escapes, $n_c^\infty/(n_c^\infty+n_d^\infty)$, as a function of the aperture width $w$. (b) Proportion of direct ($n_d(\ell)/(n_c^\infty+n_d^\infty)$) and corner ($n_c(\ell)/(n_c^\infty+n_d^\infty)$) escapes as a function of the travelled distance $l$ within a pool for experiments (filled circles) and simulations (lines, for $p_0=0.7$). Data are shown for an aperture width $w=\SI{15}{\micro\meter}$. The dashed lines represent experimental data fitted with a model of type $A\left(1-e^{-B\ell}\right)$.}
\end{figure}

This approach qualitatively reproduces the experimentally observed exponential distributions for $\ell$ as well as the decay of the characteristic travelled distance $\ell_{\rm c}$ with $w$ (Fig.~\ref{fig:Travelled distance}b). While for $p_0 = 0$ -- i.e., in the absence of \textit{corner} escapes -- the average travelled distance in the pool is significantly higher than that observed in experiments and indeed $\ell_{\rm c}$ diverges ($\ell_{\rm c}^{-1}\rightarrow 0$ for $w\rightarrow a$), we find that $p_0 = 0.70$ best-matches our experimental data. Hence, we unambiguously report that cells are biased in their trajectory by the presence of curved walls, with CR repeatedly sliding along the wall towards the exit to escape. Figure~\ref{fig:Corner escapes}a shows how the ratio of corner escapes over the total number of escapes depends on the size of the aperture, with corner escapes accounting for $\sim 90\%$ of total events for the smallest $w$. As an independent verification, our minimal model with the same $p_0 = 0.70$, and without any extra fitting parameter, once again matches the measured values well (Fig.~\ref{fig:Corner escapes}a and b).

Can we separate the narrow escape dynamics into two distinct independent -- additive -- processes, i.e., \textit{corner} and \textit{direct escapes}? Under this assumption they would be characterized by different escape rates, $\lambda_{\rm c}$ and $\lambda_{\rm d}$ respectively, leading to a simple balance of escape types:
\begin{equation}
	\begin{split}
		\dv{n_{\rm c}}{\ell} &= \lambda_{\rm c}(1-n_{\rm t})\\
		\dv{n_{\rm d}}{\ell} &= \lambda_{\rm d}(1-n_{\rm t})\\
		\dv{n_{\rm t}}{\ell} &= \lambda_{\rm t}(1-n_{\rm t}) = \left(\lambda_{\rm c}+\lambda_{\rm d}\right)(1-n_{\rm t})
	\end{split}
\end{equation}
where $n_{\rm c}(\ell)$, $n_{\rm d}(\ell)$ and $n_{\rm t}(\ell)$ denote the normalized number of corner, direct and total escapes, and $\lambda_{\rm t}$ the total escape rate. The solution to these equations were fitted to the experimental data (and compared with the results of simulations) to extract the values of the various escape rates $\lambda$ involved, as shown in Fig.~\ref{fig:Corner escapes}b for $w=15\,\mu m$.

\begin{figure}[tb]
	\includegraphics[width=1.\columnwidth]{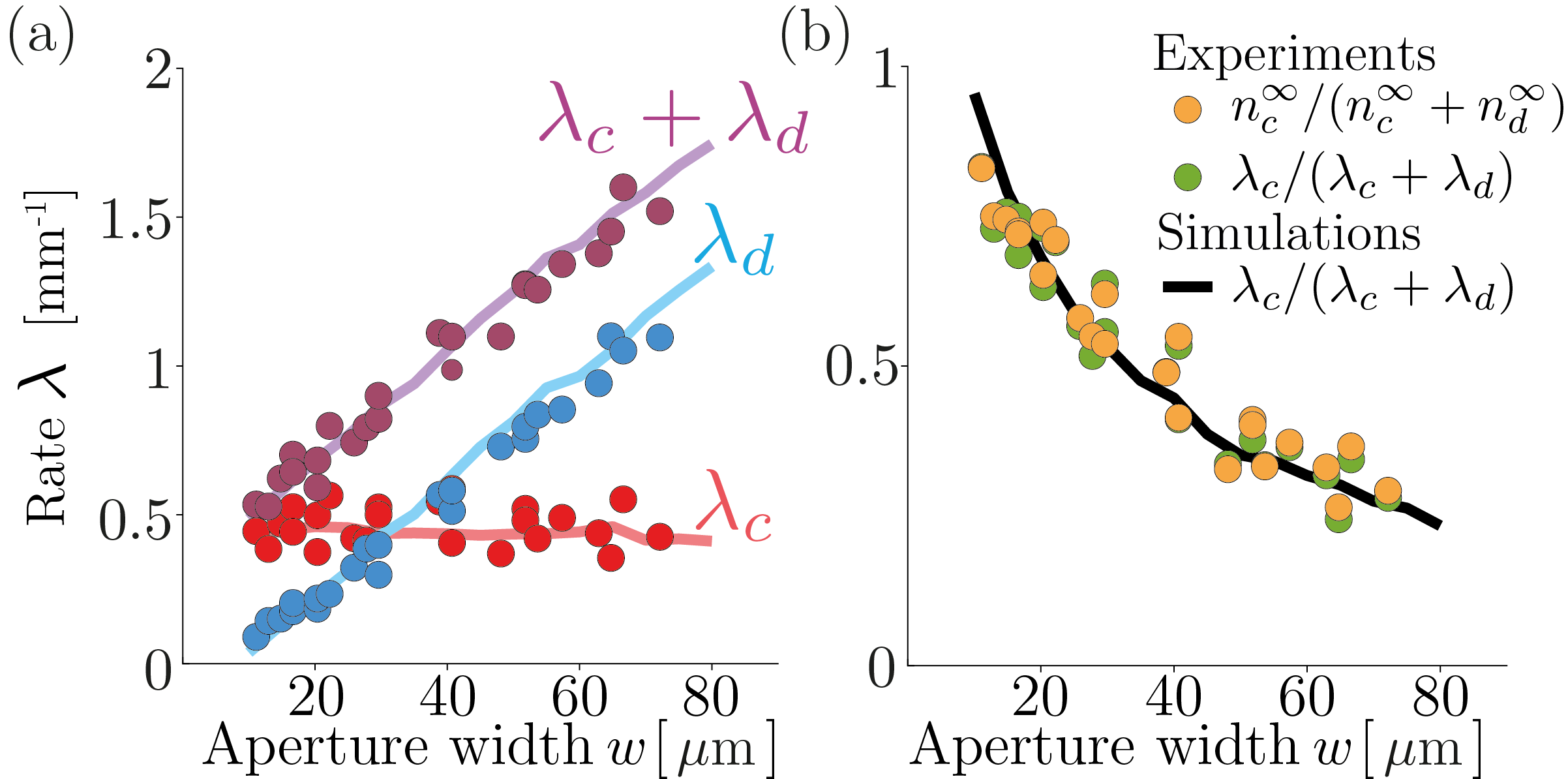}
	\caption{\label{fig:Rates}\textbf{Escape rates are additive.} (a) Corner ($\lambda_c$), direct ($\lambda_d$), and total ($\lambda_c+\lambda_d$) escape rates for various aperture sizes,from experimental (dots) and simulation (lines). (b) Comparison between the experimental proportion of corner escapes $n_c^\infty/(n_c^\infty+n_d^\infty)$ reminded from Fig.~\ref{fig:Corner escapes}a, and the ratios $\lambda_{\rm c}/\left(\lambda_{\rm c}+\lambda_{\rm d}\right)$ issued from both experiments and simulations (for $p_0=0.7$).}
\end{figure}

For all explored $w$, the rate of total escapes $\lambda_{\rm t}$ simply equals the sum of $\lambda_{\rm c}$ and $\lambda_{\rm d}$ (Fig.~\ref{fig:Rates}a). Moreover, we observe an excellent agreement between experimental (dots) and numerical (lines) results. In particular, the rate of corner escapes is constant over all aperture sizes, while the rate of direct escapes increases linearly, as expected from simple geometric arguments for the  likelihood that a hit on the boundary happens where the exit is if swimmers scatter off walls with fixed scattering angle $\theta_{\rm{ out}}$. Consequently, the proportion of corner escapes (shown in Fig.~\ref{fig:Corner escapes}a) could simply be understood as the ratio $\lambda_{\rm c}/\left(\lambda_{\rm c}+\lambda_{\rm d}\right)$ as confirmed in Fig.~\ref{fig:Rates}b and which is, once again, well captured by our numerical results.

We present the first experimental study of the narrow escape problem for micro-swimmers in circular pools. We find that for sufficiently small pools, the escape dynamics is well captured by a weakly-stochastic billiard with particles scattering off walls at fixed outgoing angles. We demonstrate that a crucial role is played by the hydrodynamic and steric forces that cause swimming cells to adhere to the pool walls and slide, guiding them to and through the exit. This results in two escape mechanisms with very distinctive trends that, when combined, elicit a significantly faster escape from confinement in a broad range of swimming parameters. The importance of boundary interactions for the escape through exits close to the active particles' size, points to the importance to investigate further the effect of micro-domain geometry in the immediate vicinity of narrow channels and the effect this has on microswimmers' dynamics. Finally, our work is neatly supported by numerical simulations that recapitulate our experimental observations. Altogether, these highlight the significance of considering the details of cell/wall interactions for the narrow escape problem of active particles.

\section{Acknowledgements}
We acknowledge financial support from grants CTM2017-83774-P and IED2019-000958-I (IT), PID2019-104232GB-I00 (IT and MP) from the Spanish Ministerio de Ciencia e Innovaci\'on (MICINN), the Ram\'on y Cajal Program (RYC-2018-02534; MP), ECOST-STSM-Request-CA17120-47203 for the COST Action (European Cooperation in Science and Technology); RPG-2018-345 (AA and MP) from The Leverhulme Trust; H2020 MSCA ITN PHYMOT (Grant agreement No 955910; IT and MP). MS also acknowledges A. Marin for his support.

\bibliography{main}

\end{document}